\PassOptionsToPackage{table}{xcolor}
\documentclass[journal]{IEEEtran}
\IEEEoverridecommandlockouts
\usepackage{times,amsmath,color,amssymb,graphicx,epsfig,cite,psfrag,subfigure,balance}
\usepackage{svg}
\usepackage{caption}

\usepackage{capt-of}
\usepackage{amsfonts,pifont,enumerate,cases}
\usepackage{mathrsfs} 
\usepackage[table]{xcolor} 
\usepackage{verbatim} 
\usepackage{bm}
\usepackage{cuted,stfloats}
\usepackage{algorithm}
\usepackage{algorithmic}

\usepackage{balance}
\usepackage{longtable}
\usepackage{blindtext}
\usepackage{multirow}
\usepackage{float}
\usepackage{threeparttable}
\usepackage{makecell}
\usepackage[utf8]{inputenc}
\usepackage{url}
\usepackage{booktabs}
\usepackage{amssymb}
\usepackage{bbding}
\usepackage{pifont}
\usepackage{wasysym}
\usepackage{utfsym}
\usepackage[algo2e,ruled,linesnumbered,lined,boxed,commentsnumbered]{algorithm2e}
\usepackage[
    colorlinks=true,
    linkcolor=blue,
    citecolor=blue,
    urlcolor=magenta
]{hyperref}

\begin{document}
\title{GML-Based Optimization for Movable Antenna Wireless Networks: Challenges and Opportunities}
\author{Zhendong Li, Yujie Zhao, Zhou Su, Tom H. Luan, Zhiqing Wei, Ying Wang, and Wen Chen 
\thanks{Zhendong Li is with the School of Information and Communication Engineering, Xi’an Jiaotong University, Xi’an 710049, China (email: lizhendong@xjtu.edu.cn). Yujie Zhao is with the School of Software, Xi’an Jiaotong University, Xi’an 710049, China (email: 2224111483@stu.xjtu.edu.cn). 
Zhou Su and Tom H. Luan are with the School of Cyber Science and Engineering, Xi'an Jiaotong University, Xi'an 710049, China (email: zhousu@ieee.org; tom.luan@xjtu.edu.cn). Zhiqing Wei and Ying Wang are with the State Key Laboratory of Networking and Switching Technology, Beijing University of Posts and Telecommunications, Beijing 100876, China (e-mail: weizhiqing@bupt.edu.cn; wangying@bupt.edu.cn). Wen Chen is with the School of Integrated Circuits, Shanghai Jiao Tong University, Shanghai 200240, China (e-mail: wenchen@sjtu.edu.cn).}\thanks{(Corresponding author: Zhou Su)}
\vspace{-1.5em}}
\maketitle
\thispagestyle{empty}


\maketitle
		
\begin{abstract}
Movable antenna (MA) is proposed as an emerging technology for future wireless networks. By leveraging the additional spatial degrees of freedom, MA can proactively reshape the wireless propagation environment, thereby enhancing network performance.
However, fully unlocking the potential of MA networks necessitates the joint optimization of MA 
antenna positioning and beamforming. 
For this non-convex and highly coupled problem, 
existing solutions exhibit significant limitations. 
Therefore, this paper develops a gradient-based meta learning (GML) 
optimization framework for MA wireless networks. 
Specifically, the framework integrates constraint handling strategies into the joint 
optimization of antenna positioning and beamforming.
We first elaborate on the hardware architecture and channel characteristics of MA, 
based on which we analyze the primary challenges in optimizing MA wireless networks. 
Subsequently, we introduce the fundamental logic of the GML framework and compare 
it with existing methods. Furthermore, we discuss the constraint handling strategies 
for applying the proposed optimization framework to MA networks.
 A specific case is studied to show the performance of the proposed framework based on numerical 
 simulation.
Finally, this paper outlines future research directions for both the GML framework and MA wireless networks.
\end{abstract}
		

\section{Introduction}

   \IEEEPARstart {W}{ith} the evolution of sixth-generation (6G) wireless networks, 
   ultra-high communication capacity and extreme reliability have become the core visions of
    development \cite{6G1}. However, existing wireless networks are predominantly based on 
    fixed position antenna (FPA). Since antenna elements in FPA are deployed at fixed positions, 
    the array geometry remains immutable once installed. 
    This fixed spatial distribution prevents the networks from dynamically adapting to 
    channel variations across space, thereby constraining the attainable spatial degrees of 
    freedom (DoFs)  and limiting network capacity and reliability. 
    To address this, movable antenna (MA)
    has recently been proposed as a novel antenna paradigm,  capable of overcoming the spatial 
    DoFs limitations of FPA \cite{MA5}. MA encompasses not only mechanically implemented structures but 
    also fluid-based structures, the latter being referred to as fluid antennas \cite{FA}.
    Unlike FPA, MA networks enable the dynamic reconfiguration of 
    array geometry according to instantaneous channel conditions\cite{MA3}, thereby enhancing 
    network performance by exploiting higher spatial DoFs.
    
    Although MA significantly improves network  performance by leveraging spatial DoFs, realizing these gains in practical systems faces severe implementation challenges. Unlike traditional FPA, MA requires precise optimization of antenna positioning within the spatial domain. This transforms communication network design from purely beamforming optimization into a more complex joint optimization problem involving both antenna positioning and beamforming. Therefore, it is crucial to develop optimization schemes that can efficiently identify favorable antenna positions from a large set of candidates while complying with practical hardware constraints. The lack of such efficient solutions has become a critical bottleneck preventing MA technology from transitioning from theoretical studies to practical deployment.

     From a mathematical perspective, MA networks involve the joint optimization of antenna 
     positioning and beamforming, which constitutes a complex non-convex problem. 
     Traditional alternating optimization (AO) algorithms based on convex optimization 
     typically rely on complex mathematical derivations and approximations \cite{MA_AO1,MA_AO2}, 
     tailored to specific objective functions and constraints. 
     Once new scenario constraints are introduced or the objective function 
     is altered, the entire algorithm often necessitates a mathematical reconstruction 
     and derivation from scratch. This lack of universality in design severely hinders 
     the flexible deployment and evolution of algorithms in diverse 6G scenarios. 
     On the other hand, regarding the optimization algorithms based on deep learning 
     that have emerged in recent years \cite{MA_learning1,MA_learning2}, they predominantly follow a data-driven learning paradigm. The trained optimizers depend heavily on the distribution of the training dataset. Consequently, when varying channel distributions occur, these methods often exhibit poor generalization and insufficient robustness, making it difficult to achieve reliable online adaptation under zero-shot or few-shot conditions.


    Given that research on MA wireless networks is still at a nascent stage 
    and in light of the aforementioned challenges, this paper investigates 
    the joint optimization of antenna positioning and beamforming in MA wireless 
    networks. Recently, gradient-based meta learning (GML) has 
    emerged as a promising optimization paradigm for non-convex problems \cite{Meta2}. 
    In particular, \cite{Meta5} develops a GML-based joint optimization algorithm in multi-waveguide 
    pinching antenna systems. 
    Centered on GML optimization logic, this paper develops a GML  
    optimization framework for MA wireless networks.
    This framework seamlessly integrates the specific constraint handling 
    mechanisms of MA networks. Unlike data-driven learning paradigms that attempt 
    to directly fit the mapping relationship for variable optimization based on 
    training data, GML adopts a ``learning to optimize" strategy. 
    Specifically, the GML algorithm takes gradient information as 
    the input to the neural network, leveraging gradients to 
    guide the variable optimization updates within the non-convex landscape\cite{Meta6}. 
    Through this meta learning mechanism, the network moves beyond rote memorization of specific mapping strategies. Instead, it learns to dynamically plan the descent trajectory based on current gradient features. This design, on the one hand, eliminates the need for extensive offline pre-training, making it a pre-training free optimization algorithm. On the other hand, since GML utilizes  the fitting capability of neural networks to determine the descent path, it does not rely on the convex geometry properties of the objective function. 
    Consequently, the proposed optimization framework provides a practical online optimization solution for MA wireless networks.

Based on the above discussions, the main contents of this paper are presented below. First, the hardware architecture and channel characteristics of MA are elaborated. Next, we introduce the fundamental principles of GML, benchmarking it against existing approaches. Subsequently, constraint-handling strategies for applying GML within MA networks are discussed. To demonstrate the efficacy of the proposed optimization framework, a specific case study is then presented. Finally, the paper concludes by outlining potential future research directions.

	\section{Architecture and \\Optimization Challenges for MA Networks}\label{sec2}


    In this section, we primarily introduce the MA hardware architecture and channel characteristics of MA wireless networks, followed by a thorough mathematical dissection of the inherent challenges in the joint optimization of antenna positioning and beamforming  problem.
    \vspace{-0.5em}
    \subsection{MA Hardware Architecture}\label{hardware}
\begin{figure}[t]  
    \centering
    \includegraphics[width=0.8\linewidth]{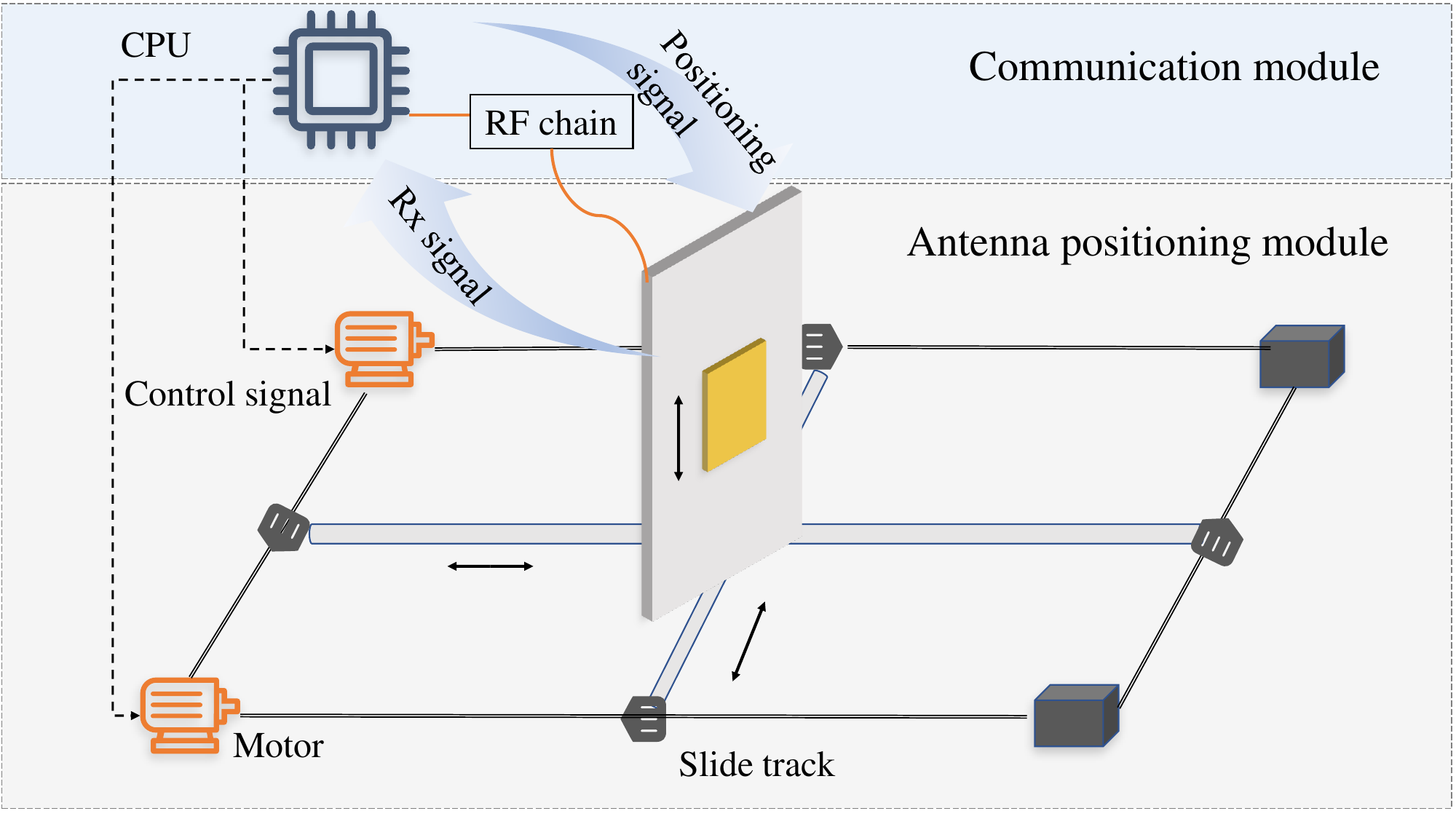} 
    \caption{MA hardware architecture.}
    \vspace{1.5em}
    \label{fig:hardware}
\end{figure}

    Unlike FPA, MA enhances network performance by actively reshaping the channel environment through dynamic antenna positioning within a specific region. Specifically, as illustrated in Fig. \ref{fig:hardware}, the MA hardware architecture consists of two main components: the communication module and the positioning module \cite{MA_hardware}.
    For the communication module, while it mirrors the architecture of conventional FPA networks, the MA is linked to the radio frequency (RF) chain using a flexible cable, thereby enabling dynamic antenna positioning.
    Antenna positioning is primarily executed by the central processing unit (CPU). Based on digital signal processing outcomes, the CPU utilizes intelligent algorithms to determine the target positions for MA elements and subsequently drives the positioning module to implement the movement. Therefore, efficient antenna positioning algorithms are crucial. In terms of the positioning module, a mechanical slider powered by stepper motors is utilized to hold the MA. Once the CPU issues a control directive, the motors operate in synchronization to shift the MA to its desired location. 

    \subsection{Channel Characteristics of MA Wireless Networks}\label{channel}
    For MA wireless networks with $M$ MA elements, as illustrated in Fig. \ref{fig:model}, whether in simple line-of-sight (LoS) scenarios or complex scattering environments, the channel response is essentially composed of steering vectors corresponding to specific directions of arrival (DoA).
    Under the far field assumption, the incident signal is modeled as a plane wave. For an antenna element located at coordinate $\mathbf{r}_m$, the phase lag of the received signal relative to the reference origin depends strictly on the geometric projection of the element's position onto the DoA. Specifically, this is mathematically expressed as the dot product of the position vector $\mathbf{r}_m$ and the wave vector $\mathbf{k}_i$\cite{MAISAC5}, where $\mathbf{k}_i$ refers to the wave vector of the $i$-th communication user. This implies that the physical movement of the antenna directly alters its interception point on the electromagnetic wavefront, thereby linearly changing the signal propagation path difference.
    Subsequently, this linear variation in path length is mapped onto a phase rotation via the factor $\text{2}\pi/\lambda$, where $\lambda$ is the carrier wavelength. As wireless networks evolve towards higher frequency bands, the value of $\text{2}\pi/\lambda$ becomes extremely large. This implies that even a minute displacement in the antenna position $\mathbf{r}_{m}$ can induce a significant phase rotation within the interval $[\text{0}, \text{2}\pi]$.
    Ultimately, this phase variation is mapped into a non-linear change within the complex exponential function $e^{-j \mathbf{k}_i^\text{T} \mathbf{r}_m}$, representing a transcendental function mapping. In other words, as the MA element is displaced, $e^{-j \mathbf{k}_i^\text{T} \mathbf{r}_m}$ rotates rapidly along the unit circle in the complex plane. This induces severe periodic oscillations in the channel, rendering the channel-based optimization objective function riddled with dense local extrema across the continuous space.
    {For channel models beyond the far field model, 
    GML can retain its gradient driven update structure as long as gradients
    with respect to antenna positions can be computed.

     \vspace{-0.5em}
     \begin{figure}[H]  
    \centering
    \includegraphics[width=0.8\linewidth]{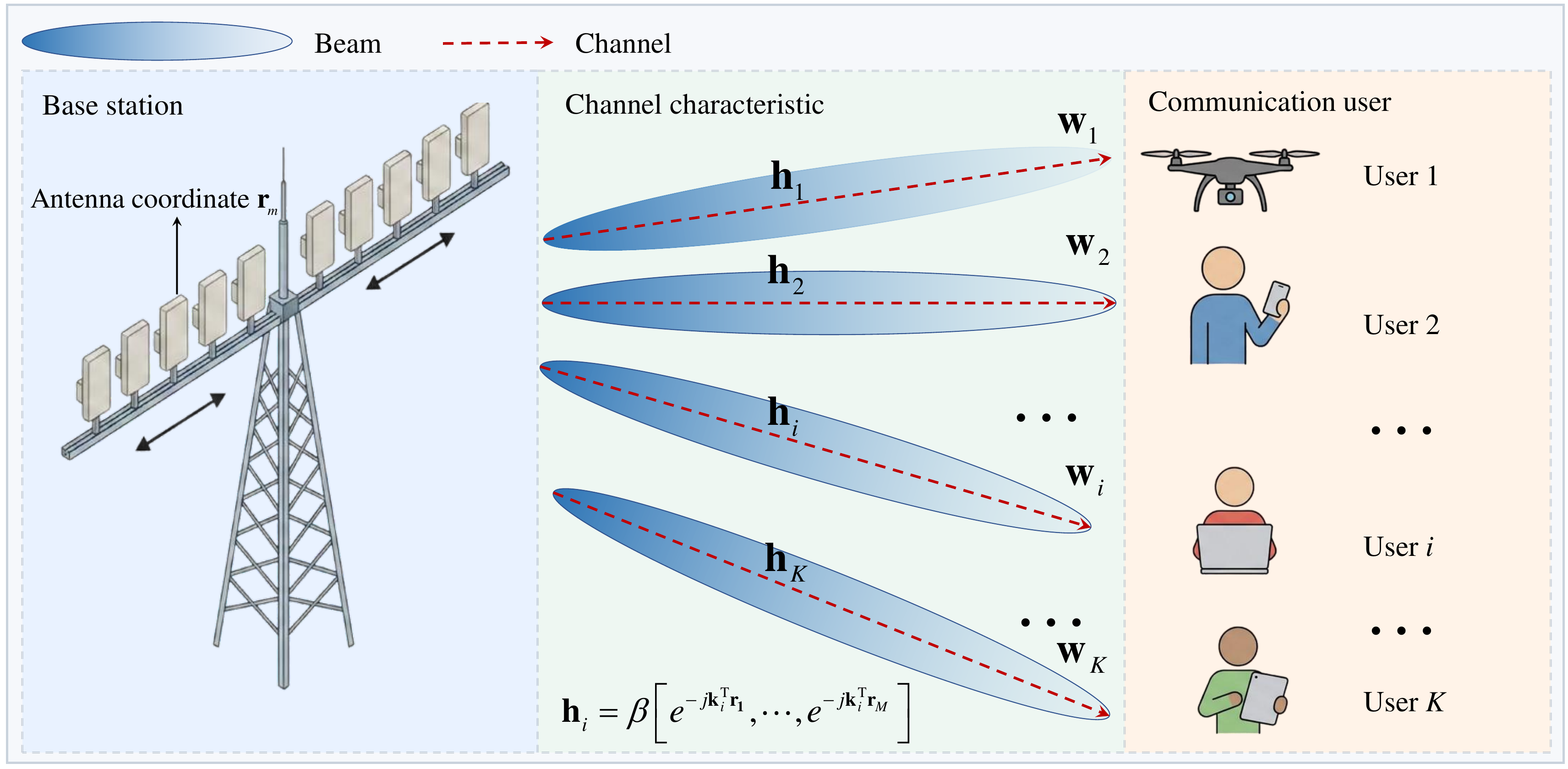} 
    \caption{MA wireless networks ($\mathbf{h}_i$ refers to the channel of the $i$-th user. $\mathbf{w}_i$ refers to the optimized beamforming vector to the $i$-th user. $\beta$ refers to the path gain).}
    \vspace{2.0em}
    \label{fig:model}
\end{figure}
    \subsection{Optimization Challenges}
    Combining the analysis of the hardware architecture and channel characteristics, the optimization of MA wireless networks is ultimately formulated as a complex mathematical problem. The complexity of this problem arises primarily from two aspects.

    First, hardware physical limitations constitute an extremely intricate non-convex feasible region. As discussed in subsection \ref{hardware}, the reconfiguration of the antenna array must be strictly confined to a specific movement region $\mathcal{C}$. Furthermore, any two elements must satisfy minimum spacing constraints to avoid physical collision and mutual coupling. Mathematically, this minimum spacing constraint not only introduces strong coupling among the $M$ position variables but also renders the feasible region itself non-convex. This implies that simple linear interpolation paths may traverse infeasible regions, thereby severely restricting the search path of algorithms. Second, field response characteristics lead to a highly multi-modal objective function. Within the aforementioned complex constraints, we must also address the non-linear mapping problem analyzed in subsection \ref{channel}. Since position variables are embedded within phase terms, the objective function (e.g., sum rate) exhibits drastic periodic oscillations within the feasible region, riddled with dense local optima traps.

    In summary, MA optimization essentially involves searching for a high quality solution of 
    a non-convex objective function over a non-convex feasible region. This non-convexity is 
    the root cause of the difficulties faced by existing algorithms. 
    Traditional numerical methods like AO struggle to directly handle non-convex constraints, 
    often relying on tedious approximation transformations. 
    Meanwhile, learning-based algorithms like deep learning fail to guarantee that 
    their output actions strictly satisfy these hard geometric constraints, 
    resulting in the generation of infeasible solutions.
    \begin{table*}[htbp] 
\centering
\caption{Comparison of GML vs. Other Optimization Algorithms for MA Networks}
\label{tab:algorithm_comparison}
\small 
\begin{tabular}{
    >{\raggedright\arraybackslash}p{0.15\textwidth} 
    >{\raggedright\arraybackslash}p{0.26\textwidth} 
    >{\raggedright\arraybackslash}p{0.26\textwidth} 
    >{\raggedright\arraybackslash}p{0.26\textwidth}
    }
\toprule
\textbf{Metrics} & \textbf{GML Algorithm} & \textbf{Deep Learning Algorithm} & \textbf{AO Algorithm} \\
\midrule
\textbf{Interpretability} & \textbf{Moderate} \newline Integrates gradient information with neural networks & \textbf{Low} \newline End-to-end mapping lacking physical insights & \textbf{High} \newline Based on explicit physical models and mathematical derivation \\
\midrule
\textbf{Computational Cost Drivers} & Repeated gradient evaluations and online neural-optimizer updates; cost depends on network size and update budget. & Forward inference after offline training; cost depends on model size, with training cost accounted for separately. & Repeated subproblem solutions; cost depends on the formulation, solver, and iteration budget.\\
\midrule
\textbf{Generalization Capability} & \textbf{Strong} \newline Learning to optimize in diverse scenarios & \textbf{Limited} \newline Performance deterioration on data outside the training set & \textbf{Poor} \newline Requires mathematical re-derivation for changing constraints \\
\midrule
\textbf{Dependence on Training Data} & \textbf{None} \newline Does not require any training data & \textbf{High} \newline Relies on datasets for pre-training & \textbf{None} \newline Does not require any training data \\
\bottomrule
\label{comparison}
\end{tabular}
\vspace{-2.5em}
\end{table*}

    \section{Gradient-Based \\Meta Learning Optimization Framework}\label{sec3}
    To address the aforementioned optimization challenges, this section introduces the GML algorithm. We provide a detailed elaboration of the GML algorithm, focusing specifically on its basic logic and distinct advantages.
    \vspace{-0.5em}
    \subsection{Basic Logic of GML}
    Following \cite{Meta2}, meta learning here refers to 
    ground level learning to optimize within a single optimization 
    problem. The inner procedure uses gradient-input neural optimizers 
    to update the position and beamforming related variables, 
    while the outer procedure adjusts the optimizer parameters 
    using the resulting global loss. 
    Thus, the update rules are learned as the coupled subproblems 
    evolve during optimization. This is the meaning of the term ``meta learning'' in this paper.

    Therefore, the basic logic of GML is to leverage the non-linear 
    approximation capability of neural networks to fit a function that 
    maps input gradient information to an output update step. 
    Specifically, the core idea of the GML algorithm is to 
    construct a dedicated neural network for each optimization variable. 
    It takes the gradient of the objective function with respect to the variable 
    as input and utilizes the network's output as the update step size. 
    Furthermore, a global loss function is formulated to govern the updating of 
    network parameters. This gradient-input  mechanism significantly 
    enhances the overall interpretability of the algorithm.
    It is widely acknowledged that in classical gradient descent, the update step of an optimization variable is defined as the product of the learning rate and the gradient of the objective function. Consequently, the update step can be viewed fundamentally as a function of the gradient. 
    According to the universal approximation theorem, neural networks are theoretically capable of approximating any continuous function. The network thus learns a gradient-to-update mapping,
whose input has an explicit optimization meaning.
    \vspace{-0.5em}
    \subsection{Advantages of GML}
  Table \ref{comparison} summarizes the key differences between the GML, 
  conventional deep learning, and AO algorithms. 
  The computational comparison identifies 
  the dominant cost factors of each paradigm, 
  while the actual runtime depends on the implementation, 
  problem size, and stopping criterion. Furthermore, applying GML to the problem of antenna positioning and beamforming in MA networks offers several distinct advantages, as detailed below:
    \subsubsection{Gradient-Input Mechanism}
   GML algorithm assigns a dedicated neural network to each optimization variable. In this architecture, the gradient of the objective function is fed into the neural network, which then outputs the requisite step size for the update. Unlike conventional methods that rely exclusively on variable-to-variable mappings, this gradient-driven approach achieves superior optimization outcomes by exploiting the  high-dimensional information embedded within the gradients. Furthermore, incorporating gradient information improves the interpretability of the framework. The neural network can be viewed as learning a mapping from gradient signals to adaptive update steps, enabling the optimizer to adjust its behavior according to the local optimization landscape.
    \subsubsection{Meta Learning Framework}
    Another significant advantage of the GML algorithm lies in its meta learning framework. Fundamentally, both traditional convex optimization methods and various machine learning or deep learning  algorithms can be conceptualized as processes that construct an optimizer for each optimization variable. For convex optimization, this construction is achieved through rigorous mathematical derivations, whereas for learning-based algorithms, the optimizer is fitted using extensive training datasets. However, once the optimizer is established and deployed to solve new problems, the parameters of the optimizers in these methods remain essentially static. Consequently, when the parameters of the problem scenario fluctuate significantly, the robustness of these algorithms may be compromised. In contrast, GML overcomes this limitation by dynamically updating the neural optimizer during the execution process. As a result, the optimizer can adapt its update strategy to the characteristics of the current scenario, enabling the algorithm to maintain robust performance even when the operational environment changes. Therefore, even when problem scenario parameters vary, GML avoids the limitations of traditional data-driven learning paradigms.
    \subsubsection{Free Pre-Training}
    Furthermore, the optimization logic of GML 
    reveals another significant advantage over traditional 
    learning-based algorithms: the elimination of extensive 
    pre-training. 
    Note that free pre-training means 
    that no offline pre-training stage is required.
    The neural optimizers are still trained online using objective 
    evaluations and gradients from the current problem.
    Conventional methods typically require an offline pre-training phase on existing datasets to tune network parameters before deployment. While this paradigm ensures performance on data following similar distributions, it suffers from two major drawbacks. First, the pre-training process often introduces high computational overhead. Second, the learned model becomes inherently dependent on the specific training dataset, which may degrade its performance when the environment changes. In contrast, GML bypasses the pre-training process entirely. The network parameters of GML are adaptively adjusted based on the immediate characteristics of the current problem scenario. This approach not only guarantees performance but also significantly reduces the computational costs associated with pre-training.
  
		\section{Application of GML \\for MA Wireless Networks}\label{sec4}
        This section discusses the major challenges of directly applying GML to MA wireless networks and presents practical strategies for handling the associated constraints.
        \vspace{-0.5em}
		\subsection{Challenges of Directly Applying GML}
        The GML framework described in section \ref{sec3} is primarily designed for unconstrained optimization problems.
        However, optimization problems in the majority of MA communication scenarios are typically formulated as constrained problems. Therefore, how to adapt the GML framework to address these constrained scenarios is a topic worthy of discussion.
        
        Constraint handling strategies can be broadly classified into two categories. The first constructs penalty terms associated with the constraints and incorporates them into the global loss function, thereby directly steering the final outputs toward the feasible domain through loss minimization. The second establishes a continuously differentiable mapping between the algorithmic output variables and the actual constrained variables. In this formulation, the neural network optimizes an unconstrained auxiliary variable, which is subsequently transformed into the target variable through the mapping function. The continuous differentiability of this mapping enables uninterrupted gradient propagation.
        Overall, these two strategies exhibit distinct advantages and limitations. Penalty based methods provide a relatively straightforward means of handling constraints. However, when multiple penalty terms are introduced, determining appropriate penalty coefficients may substantially increase the practical tuning burden of the algorithm. In contrast, mapping function based methods avoid the need to tune penalty coefficients, but continuously differentiable mappings are not always easy to construct for arbitrary constraints. Moreover, overly complex mapping functions may introduce additional computational overhead. Therefore, the choice between these two approaches should be made in light of the specific problem formulation and the characteristics of the constraints under consideration.
\begin{figure}[!t]
    \centering
    \includegraphics[width=\linewidth]{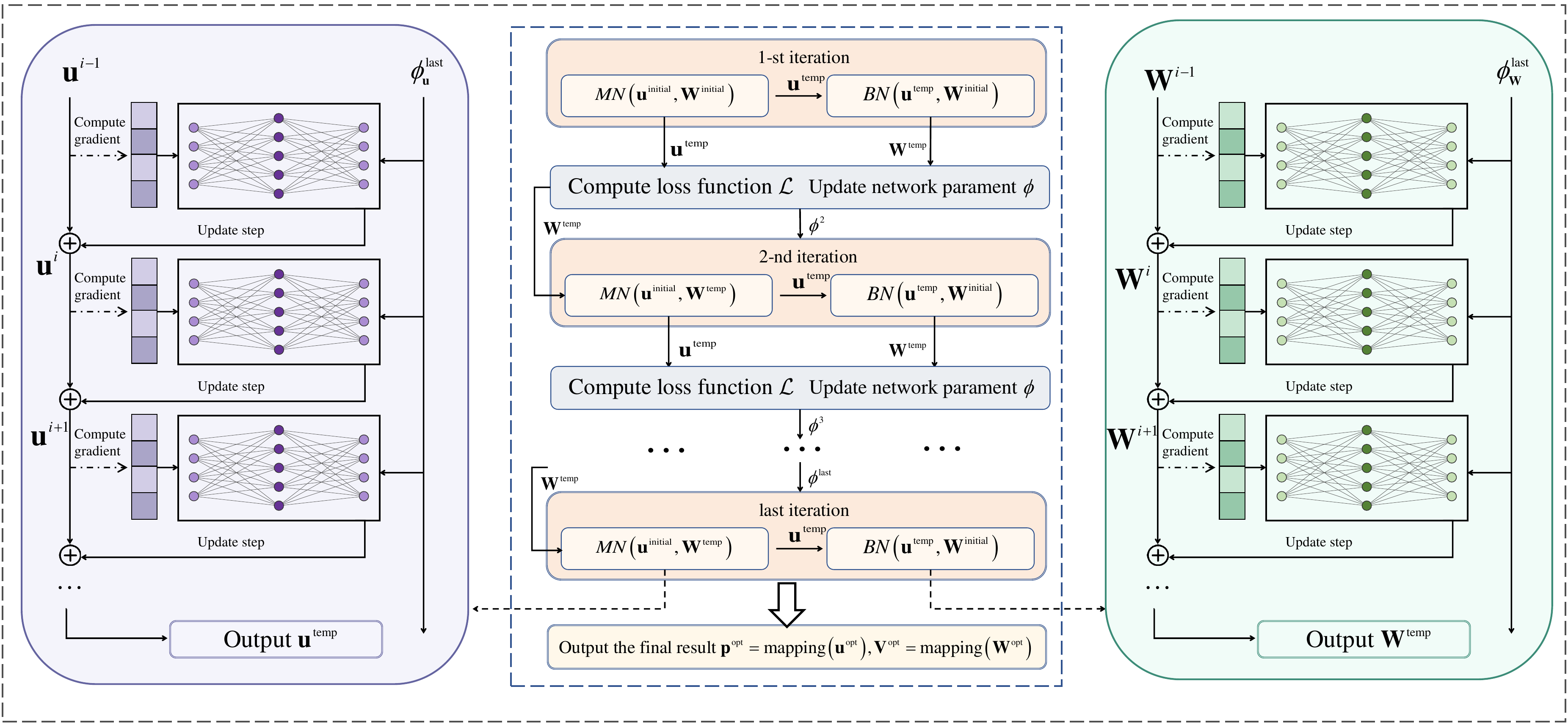}
    \caption{Flowchart of GML for MA wireless networks.
    $\mathbf{u}$ denotes the unconstrained variable obtained by applying constraint
    handling to the antenna position vector, while $\mathbf{W}$ denotes the beamforming matrix.
    $\mathbf{u}^{\text{initial}}$ denotes the initialization state of $\mathbf{u}$,
    whereas $\mathbf{u}^{\text{temp}}$ denotes its state during optimization.
    $\phi$ denotes the neural network parameters, and
    $\phi^{\text{last}} =
    \{\phi_{\mathbf{u}}^{\text{last}},
    \phi_{\mathbf{W}}^{\text{last}}\}$.}
    \label{fig:flowchart}
    \vspace{2em}
\end{figure}
\vspace{-0.5em}
        \subsection{Constraint Handling for GML for MA Wireless Networks}
        This subsection introduces strategies for handling several common constraints for MA wireless networks within the GML optimization framework, including MA position constraints, and base station (BS) power constraints.
        \subsubsection{MA Position Constraint}
        Due to the physical implementation characteristics of MA, optimization problems typically impose constraints on antenna positions. Elements must be confined within the transmit region, and the minimum spacing between any two MA elements must exceed half a wavelength. 

        For the linear MA, this paper constructs a mapping function.
        Based on the mathematical model of the linear array MA, determining the coordinates of the MA elements is equivalent to determining the intervals between adjacent elements. Considering the two endpoints of the panel, $M$ MA elements create $M+$1 intervals. Since the total length of the panel is fixed, defining these $M+\text{1}$ intervals is equivalent to defining their proportions relative to the total length. Consequently, these proportions can be modeled as an $(M+\text{1})$-dimensional vector where each element lies between 0 and 1, and their sum equals 1. The challenge then becomes constructing a mapping function that transforms an arbitrary unconstrained $(M+\text{1})$-dimensional vector into one that strictly satisfies these ratio constraints. In this paper, we employ the softmax function to map an unconstrained vector into a probability simplex serving as the ratio vector. Furthermore, to satisfy the constraint that the minimum spacing must exceed $\lambda/\text{2}$, the physical intervals are calculated by multiplying the ratios not by the total length $\text{2}L$, but by the effective free length, $\text{2}L - (M-\text{1})\lambda/\text{2}$. A fixed buffer of $\lambda/\text{2}$ is then added during the reconstruction of the absolute coordinates. This guarantees that the resulting coordinates strictly satisfy the minimum spacing constraint. Thus, we establish a continuously differentiable mapping that transforms an $(M+\text{1})$-dimensional unconstrained vector denoted as $\mathbf{u}$ into an $M$-dimensional space strictly satisfying the position constraint
        In other words, the GML algorithm is first employed to optimize these unconstrained variables, and the actual feasible MA coordinates are subsequently derived through the proposed mapping transformation. 
        
        For planar MA or MA with other architectures, 
        although constructing such a mapping function may be more complex, 
        the position constraints can still be incorporated as penalty terms, 
        thereby enabling the resulting optimization problem to be solved within the GML framework.
       Taking planar MA as an example, the gradient input and variable update mechanisms 
        remain applicable to two dimensional position variables, 
        while constraint handling must account for the array geometry. 
        For a rectangular movement region, coordinate wise differentiable mappings can 
        enforce the region boundaries, whereas penalties on pairwise distance
         violations can address minimum spacing requirements. The same applies to MA with other structures.

        \subsubsection{BS Power Constraint}
    Resource allocation problems in MA networks often involve BS power constraints, which effectively confine the transmit power within a specified threshold range. 
The BS transmit power is determined by the communication beamforming matrix $\mathbf{V}$, and the previously mentioned strategy remains applicable. Let $\mathbf{W}$ denote an unconstrained matrix. Then, the actual beamforming matrix can be obtained as $\mathbf{V}=\sqrt{P/\operatorname{Tr}(\mathbf{W}\mathbf{W}^{\mathrm{H}})}\mathbf{W}$. In this case, the GML algorithm first optimizes the unconstrained variable $\mathbf{W}$, which is subsequently mapped to obtain the actual beamforming matrix $\mathbf{V}$. Therefore, the BS power constraint can be strictly satisfied throughout the optimization process.

        \section{Case Study:\\ GML for MA Network Optimization }\label{sec5}
        Building on the preceding discussion regarding the GML optimization framework for MA networks, this section utilizes resource allocation in an MA wireless network as a case study. It demonstrates the solution approach and presents the results of applying GML to address practical resource allocation problems for MA networks.
        \begin{figure}[t] 
    \centering
    
    \includegraphics[width=0.95\linewidth]{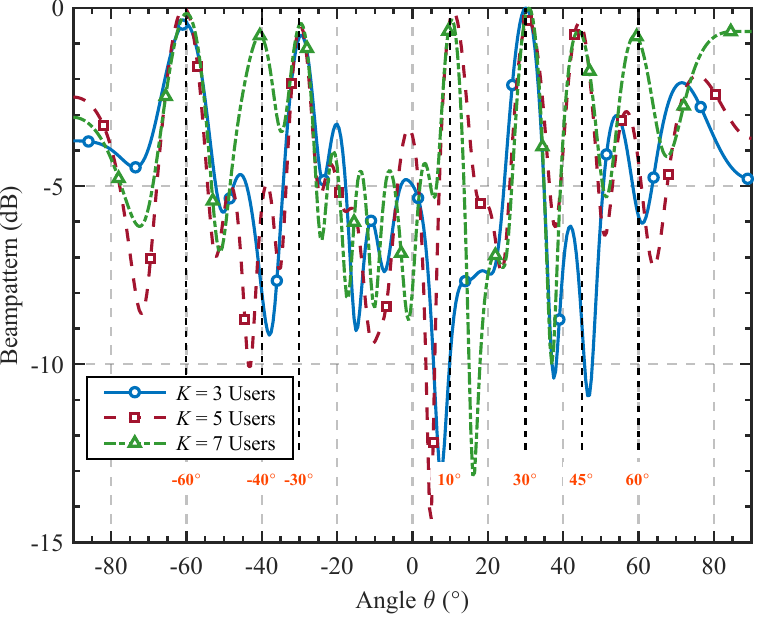} 
    \caption{Beampattern under different user distributions.}
    \label{fig:Beampattern}
    
    \vspace{4em} 
    
    \includegraphics[width=0.85\linewidth]{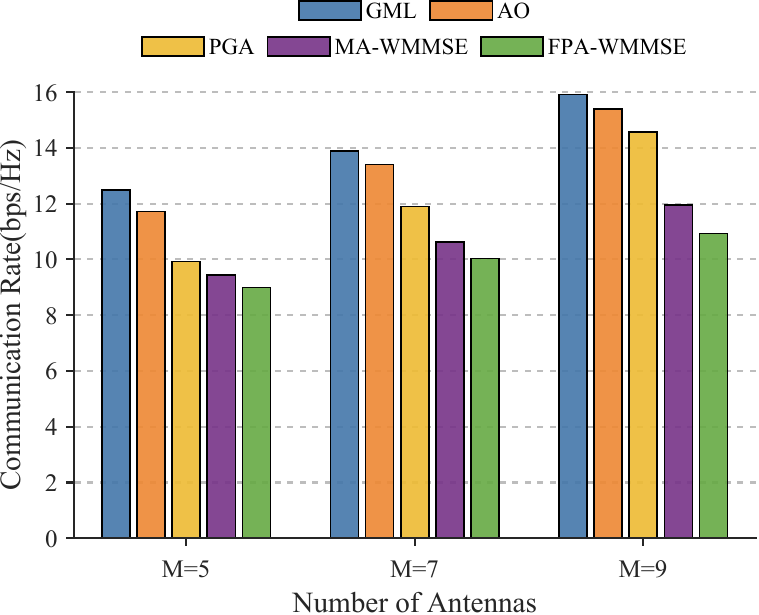} 
    \caption{Communication rate versus number of antennas.}
    \label{fig:Rate}
    
    \vspace{1.5em} 
\end{figure}

       In this case study, we formulate a problem  for an MA wireless network. Specifically, the MA BS is equipped with a linear array MA to serve users in the downlink. The objective is to maximize the network sum-rate through the joint optimization of MA antenna positioning and beamforming, subject to MA position constraints and BS power constraints.
        Due to the intricate coupling between MA positions and beamforming vectors, as well as the inherent MA position constraints, this optimization problem is non-convex. When applying the GML framework to address this problem, the MA position and BS power constraints are first processed according to the handling strategies discussed above. Subsequently, optimization iterations are performed sequentially based on the GML logic. As illustrated in Fig. \ref{fig:flowchart}, $\mathbf{p}$ and $\mathbf{V}$ denote the MA positions and the beamforming matrix, respectively. Meanwhile, $\mathbf{u}$ and $\mathbf{W}$ represent the auxiliary optimization variables derived from the constraint handling process. GML optimizes the auxiliary variables and subsequently derives the target optimization variables via a continuously differentiable mapping function. Considering the correlation between MA positions and the wireless channel, the algorithm prioritizes the optimization of MA positions. Based on the optimized positions, the wireless channel is updated, followed by the optimization of the beamforming matrix.

In the numerical simulation phase, we assume that the MA moves within a one-dimensional region. The BS transmit power is fixed at 10 dBm. Fig. \ref{fig:Beampattern} illustrates the beampattern achieved upon the convergence of the GML algorithm under different user distributions. The figure illustrates three user distribution scenarios with $\text{3}$, $\text{5}$, and $\text{7}$ users. Their corresponding azimuth angles are located at $\{\pm \text{30}^\circ, -\text{60}^\circ\}$, $\{\pm \text{30}^\circ, -\text{60}^\circ, \text{10}^\circ, \text{45}^\circ\}$, and $\{\pm \text{30}^\circ, \pm\text{60}^\circ, -\text{40}^\circ, \text{10}^\circ, \text{45}^\circ\}$, respectively. The simulation results demonstrate that through the proposed joint optimization, the network is able to precisely focus the transmit power toward the desired angles under different user distributions. As shown in Fig. \ref{fig:Beampattern}, the generated beampattern not only achieves maximum array gain in the target directions but also effectively suppresses sidelobe levels in non-target regions, thereby guaranteeing the communication performance of the network.

Fig. \ref{fig:Rate} illustrates the performance comparison between the proposed 
GML algorithm and other baseline algorithms against the number of antennas. 
All the experimental results are averaged over 50 independent channel realizations. Specifically, for the neural networks 
employed in GML, we construct a simple multilayer perceptron with one hidden layer for 
each optimization variable.
The considered baseline algorithms are as follows:
\begin{itemize}
    \item {\bf{Baseline 1}} (AO): Alternately optimizes antenna positioning and transmit beamforming by solving first order convex approximated subproblems with CVX.
    \item {\bf{Baseline 2}} (PGA): Projected gradient ascent (PGA) \cite{MASecure} is employed to optimize the MA part.
    \item {\bf{Baseline 3}} (MA-WMMSE): PGA is employed to optimize the MA part and WMMSE \cite{WMMSE} is used to optimize the beamforming part.
    \item {\bf{Baseline 4}} (FPA-WMMSE): The dual-functional BS employs an FPA array and WMMSE is used to optimize the beamforming part.
\end{itemize}
As observed in Fig. \ref{fig:Rate}, the four MA optimization algorithms (GML, AO, PGA and MA-WMMSE) significantly outperform the FPA scheme (FPA-WMMSE) across varying numbers of antennas. This underscores the substantial performance enhancement attributed to the additional DoFs inherent in MA networks. Furthermore, among the four MA optimization algorithms, GML consistently achieves superior performance compared to others, which validates the effectiveness of the proposed algorithm and the rationality of the adopted constraint handling strategy.

\section{Open Challenges and Future Directions}\label{sec6}
In this section, we propose four key future directions from the perspectives of GML algorithms and MA networks: dynamic optimization, robustness against imperfect channel state information (CSI), hardware impairments and energy efficiency, as well as synergy with integrated sensing and communication (ISAC).
\vspace{-0.5em}
\subsection{Dynamic Optimization}
Most existing MA optimization studies solve for the optimal antenna 
configuration independently within each time slot, assuming fixed user positions.
In downlink transmission, however, users may form highly dynamic 
spatial patterns, such as the gathering and dispersal of crowds. Optimization based on a 
fixed user distribution is therefore inherently short sighted, as it neglects the temporal 
evolution of user locations and may respond poorly to sudden traffic surges in hotspot areas.
 Future research must transcend this 
assumption of static user distribution and shift towards optimization 
for dynamic user distributions. This necessitates algorithms equipped with 
capabilities to sense and predict downlink user variations. 
Rather than focusing solely on instantaneous user positions, 
algorithms should reconstruct beamforming and antenna positioning 
in real time according to user distributions to effectively cover 
time-varying user regions.
\vspace{-0.5em}
\subsection{Robustness Against Imperfect CSI}
Although gradient-based optimization is theoretically efficient, 
its implementation in practical systems relies heavily on accurate CSI. 
Pilot contamination and quantization errors in feedback can 
introduce substantial noise and bias into the estimated physical gradients, 
making direct gradient-based updates potentially detrimental to system performance. 
Therefore, future research should strive to enhance the robustness of the 
GML framework against imperfect CSI. A possible direction is to develop a 
noise resilient optimization scheme using adversarial perturbations or statistical error models. 
The neural network can combine gradient inputs with historical information to detect and correct 
unreliable update directions, thereby sustaining stable performance in the presence of imperfect CSI.
\vspace{-0.5em}
\subsection{Hardware Impairments and Energy Efficiency}
Although existing studies have validated the theoretical superiority of MA networks over traditional ones, 
practical deployment confronts significant bottlenecks arising from hardware impairments. 
Continuous actuation of the electromechanical positioning system can impose 
considerable power consumption, while repeated bending of the flexible RF cables may cause 
noticeable transmission loss. Excessive antenna movement driven solely by spectral efficiency 
maximization could therefore offset the resulting communication gains and reduce the overall energy 
efficiency of the system.
Therefore, future research should establish a comprehensive energy efficiency model that accounts for both mechanical energy consumption and circuit losses, exploring the optimal trade-off between performance and energy consumption in practical systems. Benefiting from the advantage of requiring no mathematical reformulation, the proposed GML optimization framework can efficiently learn the movement policy by incorporating an energy consumption penalty term into the loss function. Consequently, it flexibly achieves this trade-off under complex physical constraints.
\vspace{-0.5em}
\subsection{Synergy with ISAC}
MA should not be regarded as an isolated technology but rather as an integral component 
of the 6G physical layer. In particular, the deep integration of MA with ISAC holds 
immense potential. Traditional ISAC systems, constrained by the static geometry of FPA, 
often suffer from uncovered sensing blind spots. Furthermore, the high correlation of target echoes 
at specific angles limits estimation accuracy. The fusion of MA and ISAC primarily addresses 
these inherent geometric limitations of FPA. By introducing actively reconfigurable antenna arrays, 
MA enables transceivers to dynamically optimize observation angles. These additional DoFs 
facilitate the simultaneous enhancement of both communication and sensing performance. 
Therefore, future research should focus on the technical synergy between MA and ISAC to further 
unlock the full potential of ISAC. 
In particular, sensing requirements in ISAC may be handled using a 
constraint treatment approach similar to that presented in this work. One possible implementation is 
to penalize the discrepancy between the designed and desired covariance matrices, 
thereby incorporating the sensing requirements into the GML objective and obtaining effective 
solutions for MA-enabled ISAC systems.

\section{Conclusion}\label{sec7}

This paper investigated optimization problems for MA wireless networks. While MA can enhance spectral efficiency by leveraging continuous spatial DoFs, it also introduces specific optimization challenges. To this end, this paper first analyzed the mathematical intractability of the joint optimization problem for MA wireless networks. We then proposed an optimization framework based on GML and compared it with existing numerical optimization and data-driven learning methods in terms of flexibility and generalizability. Furthermore, we discussed the application of the GML framework to MA networks by incorporating specific constraint handling strategies, and validated its effectiveness through a case study. Finally, we identified key challenges associated with GML and MA networks and outlined potential future directions.

\section{Acknowledgment}
This work was supported in part by NSFC under Grant 62401448, Grant 62531015, Grant U24A20237; in part by China National Post-Doctoral Program for Innovation Talents under Grant BX20240277.

\bibliographystyle{IEEEtran}
\bibliography{reference}

@article{6G1,
  author={Wang, Cheng-Xiang and You, Xiaohu and Gao, Xiqi and Zhu, Xiuming and Li, Zixin and Zhang, Chuan and Wang, Haiming and Huang, Yongming and Chen, Yunfei and Haas, Harald and Thompson, John S. and Larsson, Erik G. and Di Renzo, Marco and Tong, Wen and Zhu, Peiying and Shen, Xuemin and Poor, H. Vincent and Hanzo, Lajos},
  journal={IEEE Commun. Surveys \& Tuts.}, 
  title={On the road to {6G}: Visions, requirements, key technologies, and testbeds}, 
  year={2023},
  month={Feb.},
  volume={25},
  number={2},
  pages={905-974},
  doi={10.1109/COMST.2023.3249835}
}

@article{FA,
  author={Wong, Kai-Kit and Shojaeifard, Arman and Tong, Kin-Fai and Zhang, Yangyang},
  journal={IEEE Trans. Wireless Commun.}, 
  title={Fluid antenna systems}, 
  year={2021},
  month={Mar.},
  volume={20},
  number={3},
  pages={1950-1962},
  doi={10.1109/TWC.2020.3037595}
}

@article{MA5,
  author={Zhu, Lipeng and Ma, Wenyan and Mei, Weidong and Zeng, Yong and Wu, Qingqing and Ning, Boyu and Xiao, Zhenyu and Shao, Xiaodan and Zhang, Jun and Zhang, Rui},
  journal={IEEE Commun. Surveys \& Tuts.}, 
  title={A tutorial on movable antennas for wireless networks}, 
  year={2026},
  month={Feb.},
  volume={28},
  pages={3002-3054},
  doi={10.1109/COMST.2025.3546373}
}

@ARTICLE{MA3,
  author={Shao, Xiaodan and Mei, Weidong and You, Changsheng and Wu, Qingqing and Zheng, Beixiong and Wang, Cheng-Xiang and Li, Junling and Zhang, Rui and Schober, Robert and Zhu, Lipeng and Zhuang, Weihua and Shen, Xuemin},
  journal={IEEE Commun. Surveys \& Tuts.}, 
  title={A Tutorial on Six-Dimensional Movable Antenna for {6G} Networks: Synergizing Positionable and Rotatable Antennas}, 
  month={Aug.},
  year={2025},
  volume={28},
  number={},
  pages={3666-3709},
  doi={10.1109/COMST.2025.3602939}}

@article{MAISAC5,
  author={Li, Zhendong and Ba, Jianle and Su, Zhou and Peng, Haixia and Wang, Yuntao and Chen, Wen and Wu, Qingqing},
  journal={IEEE Trans. Wireless Commun.}, 
  title={Joint discrete antenna positioning and beamforming optimization in movable antenna enabled full-duplex {ISAC} networks}, 
  year={2025},
  month={May},
  volume={25},
  pages={7220-7234},
  doi={10.1109/TWC.2025.3630154}
}

@article{MASecure,
  author={Hu, Guojie and Wu, Qingqing and Xu, Kui and Si, Jiangbo and Al-Dhahir, Naofal},
  journal={IEEE Signal Process. Lett.},
  title={Secure wireless communication via movable-antenna array},
  year={2024},
  month={Jan.},
  volume={31},
  pages={516-520},
  doi={10.1109/LSP.2024.3359894}
}

@article{WMMSE,
  author={Shi, Qingjiang and Razaviyayn, Meisam and Luo, Zhi-Quan and He, Chen},
  journal={IEEE Trans. Signal Process.},
  title={An iteratively weighted {MMSE} approach to distributed sum-utility maximization for a {MIMO} interfering broadcast channel},
  year={2011},
  month={Sep.},
  volume={59},
  number={9},
  pages={4331-4340},
  doi={10.1109/TSP.2011.2147784}
}

@article{Meta2,
  author={Xia, Jing-Yuan and Li, Shengxi and Huang, Jun-Jie and Yang, Zhixiong and Jaimoukha, Imad M. and Gündüz, Deniz},
  journal={IEEE Trans. Neural Netw. Learning Syst.},
  title={Metalearning-based alternating minimization algorithm for nonconvex optimization},
  year={2023},
  month={Sep.},
  volume={34},
  number={9},
  pages={5366-5380},
  doi={10.1109/TNNLS.2022.3165627}
}

@article{Meta5,
  author={Zhou, Kang and Zhou, Weixi and Cai, Donghong and Lei, Xianfu and Xu, Yanqing and Ding, Zhiguo and Fan, Pingzhi},
  journal={IEEE Trans. Commun.},
  title={A gradient meta-learning joint optimization for beamforming and antenna position in pinching-antenna systems},
  month={Nov.},
  year={2025},
  volume={74},
  pages={1099-1112},
  doi={10.1109/TCOMM.2025.3634205}
}

@article{Meta6,
  author={Amhaz, Ali and Khisa, Shreya and Elhattab, Mohamed and Assi, Chadi and Sharafeddine, Sanaa},
  journal={IEEE Trans. Commun.}, 
  title={Enhancing {CoMP}-{RSMA} performance with movable antennas: A meta-learning optimization framework}, 
  year={2026},
  month={Jan.},
  volume={74},
  pages={3802-3813},
  doi={10.1109/TCOMM.2026.3657455}
}

@article{MA_hardware,
  author={Zhu, Lipeng and Ma, Wenyan and Zhang, Rui},
  journal={IEEE Commun. Mag.},
  title={Movable antennas for wireless communication: Opportunities and challenges},
  year={2024},
  month={Jun.},
  volume={62},
  number={6},
  pages={114-120},
  doi={10.1109/MCOM.001.2300212}
}

@article{MA_AO1,
  author={Tang, Jun and Pan, Cunhua and Zhang, Yang and Ren, Hong and Wang, Kezhi},
  journal={IEEE Trans. Commun.},
  title={Secure {MIMO} communication relying on movable antennas},
  year={2025},
  month={Apr.},
  volume={73},
  number={4},
  pages={2159-2175},
  doi={10.1109/TCOMM.2024.3465369}
}

@article{MA_AO2,
  author={Hu, Guojie and Wu, Qingqing and Xu, Donghui and Xu, Kui and Si, Jiangbo and Cai, Yunlong and Al-Dhahir, Naofal},
  journal={IEEE Trans. Mobile Comput.},
  title={Movable antennas-assisted secure transmission without eavesdroppers’ instantaneous {CSI}},
  year={2024},
  month={Dec.},
  volume={23},
  number={12},
  pages={14263-14279},
  doi={10.1109/TMC.2024.3438795}
}

@article{MA_learning1,
  author={Tang, Xiao and Jiang, Yudan and Liu, Jinxin and Du, Qinghe and Niyato, Dusit and Han, Zhu},
  journal={IEEE Trans. Veh. Technol.}, 
  title={Deep learning-assisted jamming mitigation with movable antenna array}, 
  year={2025},
  month={Sep.},
  volume={74},
  number={9},
  pages={14865-14870},
  doi={10.1109/TVT.2025.3558595}
}

@article{MA_learning2,
  author={Jang, Suhwan and Lee, Chungyong},
  journal={IEEE Trans. Wireless Commun.}, 
  title={Deep learning-driven channel estimation for movable antenna-aided wideband systems}, 
  year={2025},
  month={Nov.},
  volume={25},
  pages={6954-6969},
  doi={10.1109/TWC.2025.3627684}
}

\vspace{1em}
{\fontsize{8pt}{10pt}\selectfont

\noindent\textbf{Zhendong Li} is currently an Assistant Professor with the School of Information and Communication Engineering in Xi’an Jiaotong University. His research interests include movable antennas, ISAC, and wireless AI.

\noindent\textbf{Yujie Zhao} obtained the bachelor's degree from the School of Information and Communication Engineering of Xi'an Jiaotong University, and is currently pursuing the master's degree in the School of Software Engineering, Xi'an Jiao Tong University. His research interests include movable antennas, ISAC, and wireless AI.

\noindent\textbf{Zhou Su} is currently a professor with the School of Cyber Science and Engineering in Xi’an Jiaotong University. His research interests include AI security, embodied intelligence, and communication network security.

\noindent\textbf{Tom H. Luan} is currently a professor with the School of Cyber Science and Engineering in Xi’an Jiaotong University. His research interests include digital twin and game theory.

\noindent\textbf{Zhiqing Wei} is currently a professor with the School of Information and Communication Engineering, Beijing University of Posts and Telecommunications. His research interests include ISAC and intelligent machine communication and networking.

\noindent\textbf{Ying Wang} is currently a professor with the School of Information and Communication Engineering, Beijing University of Posts and Telecommunications. Her research interests include wireless resource management and network performance analysis.

\noindent\textbf{Wen Chen} is currently a professor with the School of Integrated Circuits, Shanghai Jiao Tong University. His research interests include multiple access, wireless AI, and RIS.

\par
}

\end{document}